\documentclass[12pt,preprint,notoc,nohyper]{AKRSX} 

\usepackage{epsfig,multicol,bbm}

\newcommand\fverb{\setbox\pippobox=\hbox\bgroup\verb}
\newcommand\fverbdo{\egroup\medskip\noindent%
            \fbox{\unhbox\pippobox}\ }
\newcommand\fverbit{\egroup\item[\fbox{\unhbox\pippobox}]}
\newbox\pippobox

\title{Symmetries of the Weyl tensor in Bianchi V spacetimes}

\author{A. R. Kashif$^1$\thanks{\emph{On leave from:}
College of EME, National University of Sciences and Technology,
Rawalpindi, Pakistan.}~, K. Saifullah$^2$\footnote{\emph{On leave
from:} Centre for Advanced Mathematics and Physics, National
University of Sciences and Technology, Rawalpindi, Pakistan,
\emph{and} Department of Mathematics, Quaid-i-Azam University,
Islamabad, Pakistan.} ~ and
G. Shabbir$^3$ \\

$^1$Department of Mathematical Sciences, University of Aberdeen,
Aberdeen AB24 3UE, Scotland, UK \\

$^2$School of Mathematical Sciences, Queen Mary, University of
London, London,
UK \\
$^3$Faculty of Engineering Sciences, GIK Institute of Engineering
Sciences and Technology, Topi. Swabi, NWFP, Pakistan \\
Electronic address: \email{kashmology@yahoo.com},
\email{saifullah@qau.edu.pk}, \email{shabbir@giki.edu.pk}}

\preprint{}  

\abstract{Symmetries of geometrical and physical quantities in
general relativity provide important information about the curvature
structure of the spacetimes. Symmetries of the curvature and the
Weyl tensors, known as curvature and Weyl collineations
respectively, are two of such important symmetries. Some results on
these symmetries for Bianchi type V spacetimes are discussed.}

\dedicated{}

\begin{document}

Symmetries of tensors in general relativity -- Killing vectors and
collineations -- play an important role in understanding not only
the geometric structure of the underlying spaces but their physical
properties as well. They have been used in finding new solutions of
Einstein's Field Equations (EFEs), classifying these solutions and
by virtue of Noether's theorem constructing the conservation laws
for the spacetime. Thus, invariance under the Lie transport of the
metric, Ricci and energy momentum tensors define Killing vectors
(KVs), Ricci collineations (RCs) and matter collineations (MCs),
respectively \cite{2r,9r}. Curvature collineations (CCs) which are
symmetries of the Riemann tensor are significant for studying the
curvature structure of spacetimes \cite{kit,9r}. The Weyl tensor,
$\mathbf{C}$, is fundamental in understanding the purely
gravitational field for a spacetime with the matter content removed
\cite{PR}. Its local symmetries, Weyl collineations (WCs) \cite{kit,
ibrar, GS-ARK, gs2}, are of particular interest since it is
conformally invariant \cite{lss}. Mathematically, WCs are given by
\begin{equation}
\pounds _{_{\mathbf{X}}}\mathbf{C}=0,  \nonumber
\end{equation}
where $\pounds _{_{\mathbf{X}}}$ is the Lie derivative along the
vector field $\mathbf{X}$. In component form this becomes
\[
C_{bcd,f}^{a}X^{f}+C_{fcd}^{a}X_{,b}^{f}+C_{bfd}^{a}X_{,c}^{f}+
C_{bcf}^{a}X_{,d}^{f}-C_{bcd}^{f}X_{,f}^{a}=0,
\]
where comma denotes the partial derivative. This is a system of 20
nonlinear partial differential equations as compared to the
collineations of rank two tensors (RCs and MCs, for example) which
are systems of 10 equations.

On account of its algebraic symmetries we can write the 4th rank
Weyl tensor (and the curvature tensor) of 4-dimensions in the form
of a 6 dimensional matrix, whose rank gives the rank of the tensor.
Further, while the metric tensor cannot be degenerate, the other
tensors can be and hence give way to the possibility of infinite
degrees of freedom (i.e. infinite dimensional Lie algebras) as well.
The KVs of a space form a subset of all other collineations but the
inclusion relationship between the symmetries of two fourth rank
tensors, CCs and WCs, when both are finite is yet to be established.
While it is known \cite{ibrar} that CCs can be properly contained in
WCs when both are finite, no spacetime is known to the present
authors which admits CCs which are not WCs and yet both are finite.
On the other hand, there is no proof available that this is not
possible. The Schwarzschild interior spacetime, for example, is
Petrov type O \cite{2r} and thus every vector field is a WC while
CCs are finite. The Reissner-Nordstrom spacetime is of Petrov type D
and both the WCs and CCs are finite and equal. But when we take
pressure as constant in the Schwarzschild interior we see that the
WCs are properly contained in infinitely many CCs. For vacuum
spacetimes with zero cosmological term, however,  the Ricci tensor,
$\mathbf{R,}$ is zero and WCs and CCs coincide because the Weyl
tensor reduces to the curvature tensor.

Enumeration of all Lie groups is useful in mathematics as well as in
physics. The $G_3$, for example, were originally enumerated by
Bianchi which were divided into nine types, \emph{Bianchi I} to
\emph{Bianchi IX} \cite{2r}.

Let us consider the Bianchi type V spacetimes which admit three KVs
given by \cite{2r}

\begin{eqnarray}
K^{1} &=&\frac{\partial }{\partial y} \ ,  \nonumber \\
K^{2} &=&\frac{\partial }{\partial z} \ ,   \nonumber  \\
K^{3} &=&\frac{\partial }{\partial x}-y \frac{\partial }{\partial
y}-z\frac{\partial }{\partial z}\ . \nonumber
\end{eqnarray}

These spacetimes in $(t,x,y,z,)$ coordinates can be written as

\begin{equation}
ds^{2}=-dt^{2}+A(t)^2dx^{2}+B(t)^2dy^2+(C(t)^{2}+x^{2}B(t)^{2})dz^{2}-2xB(t)^{2}dydz
. \nonumber
\end{equation}

Following a well known procedure \cite{gs1, shh} the components
$C_{abcd}$ of the Weyl tensor for these spacetimes can be written as
the $6\times6$ matrix

\begin{eqnarray}
C_{abcd}=\left(
           \begin{array}{cccccc}
             C_{1010} & 0 & 0 & 0 & 0 & C_{1023} \\
             0 & C_{2020} & C_{2030} & 0 & C_{1320} & 0 \\
             0 & C_{2030} & C_{3030} & C_{1230} & C_{1330} & 0 \\
             0 & 0 & C_{1230} & C_{1212} & C_{1213} & 0 \\
             0 & C_{1320} & C_{1330} & C_{1213} & C_{1313} & 0 \\
             C_{1023} & 0 & 0 & 0 & 0 & C_{2323} \\  \nonumber
           \end{array}
         \right) .
\end{eqnarray}

Similarly the curvature tensor can also be written as a $6\times6$
matrix.  If its rank is greater than or equal to 4 then the Lie
algebra of CCs is finite dimensional \cite{Ahb-Ark-Aq}. Now, the
rank of the Weyl matrix is always even and if it is 6 or 4 the Weyl
symmetry trivially reduces to the conformal symmetry \cite{gs1}.
Further, we note that \cite{GS-ARK} it cannot have rank 2 . Thus we
conclude that these spacetimes do not admit non-trivial WCs.

\acknowledgments

ARK and KS acknowledge a research grant from the Higher Education
Commission of Pakistan. They are also thankful to the National
University of Sciences and Technology, Pakistan for travel support
to participate in MG11, Berlin, 2006.

\end{document}